\documentclass[aps,prl,reprint,tightenlines,superscriptaddress,amsmath,amssymb]{revtex4-1} 
\usepackage{amsfonts,amsmath,amssymb}

\usepackage{accents}
\usepackage{graphicx}
\usepackage[latin1]{inputenc}
\usepackage{mathrsfs}
\usepackage{xcolor}
\usepackage{float}
\usepackage[export]{adjustbox}[2011/08/13]
\usepackage{cancel}
\usepackage{caption}
\usepackage[percent]{overpic}
\usepackage{ulem}

\def\pd{\partial}
\def\reals{{\mathbb R}}

\newcommand*{\be}{\begin{equation}}
\newcommand*{\ee}{\end{equation}}
\newcommand*{\bse}{\begin{subequations}}
\newcommand*{\ese}{\end{subequations}}
\newcommand*{\bme}{\begin{multiequations}}
\newcommand*{\eme}{\end{multiequations}}

\newcommand {\bal}{\begin{align}}
\newcommand {\eal}{\end{align}}

\renewcommand*{\pd}{\partial}
\newcommand*{\fb}{\frac}

\renewcommand{\Delta}{\triangle}

\newcommand*{\rir}{\right \rangle}
\newcommand*{\lel}{\left \langle}

\newcommand{\B}{{\bf B}}
\newcommand{\Bt}{\tilde{{\bf B}}}
\renewcommand{\u}{{\bf u}}
\newcommand{\Rm}{\text{Rm}}

\newcommand {\tcb}{\textcolor{black}}
 
\makeatletter
\def\mathcolor#1#{\@mathcolor{#1}}
\def\@mathcolor#1#2#3{%
  \protect\leavevmode
  \begingroup
    \color#1{#2}#3%
  \endgroup
}
\makeatother

\begin{document}

\title{Enhanced dynamo growth in nonhomogeneous conducting fluids}
\author{Florence Marcotte}
\email{florence.marcotte@inria.fr}
\affiliation{Inria Sophia Antipolis M\'editerran\'ee, Universit\'e C\^ote d'Azur, Inria, CNRS, LJAD, France}
\author{Basile Gallet}
\affiliation{Universit\'e Paris-Saclay, CNRS, CEA, Service de Physique de l'Etat Condens\'e, France}
\author{Fran\c cois Pétrélis}
\affiliation{Laboratoire de Physique de l'Ecole Normale Sup\'erieure, ENS, Universit\'e PSL, CNRS, Sorbonne Universit\'e, Universit\'e Paris-Diderot, France}
\author{Christophe Gissinger}
\affiliation{Laboratoire de Physique de l'Ecole Normale Sup\'erieure, CNRS, PSL Research University, France}
\affiliation{Institut Universitaire de France (IUF)}

\begin{abstract}
We address magnetic-field generation by dynamo action in systems with inhomogeneous electrical conductivity and magnetic permeability. More specifically, we first show that the Taylor-Couette kinematic dynamo undergoes a drastic reduction of its stability threshold when a (zero-mean) modulation of the fluid's electrical conductivity or magnetic permeability is introduced. These results are obtained outside the mean-field regime, for which this effect was initially proposed. Beyond this illustrative example, we extend a duality argument put forward by Favier \& Proctor {(2013)} to show that swapping the distributions of conductivity and permeability and changing ${  u}\to -{  u}$ leaves the dynamo threshold unchanged. This allows one to make connections between {\it a priori} unrelated dynamo studies. Finally, we discuss the possibility of observing such an effect both in laboratory and astrophysical settings.
\end{abstract}

\maketitle

\section{Introduction}

Temperature or concentration gradients in astrophysical flows play a critical role in magnetogenesis, driving buoyant overturning and generally constraining the structure and intensity of fluid motions. These motions can in turn power a dynamo instability in electrically conducting fluids when they are morphologically favorable and vigorous enough to overcome ohmic dissipation. This results in the amplification and persistence of long-lived, coherent magnetic fields \cite{L19}. However, spatial variations in temperature, density or concentration of chemical species not only govern the flows' dynamics but - importantly - they also induce spatial variations of the electrical conductivity, which can reach considerable amplitudes in astrophysical plasmas \cite{S62}.

The combined effect of variable electrical conductivity and density stratification on magnetic field morphology has been investigated numerically for specific models of giant planets  \cite{D13,DJ18}. Moreover, it has been shown that a spatially inhomogeneous conductivity can provide the condition for a dynamo instability to build up in flow configurations and regimes where no magnetic fields could be amplified otherwise: by considering the interaction of magnetic field and conductivity fluctuations over a spatial scale small compared to that of the mean magnetic field, \cite{PAG16} recently demonstrated the existence of a new mean-field amplification mechanism, which they refer to as $\alpha^{\sigma}$-effect in reference to the classical $\alpha$-effect associated with velocity fluctuations \cite{HKM}. They argued that such conductivity variations could account for the magnetic field morphology of Neptune and Uranus, and this model has been applied to also investigate possible dynamo action in Jupiter atmosphere \cite{RME17}.

Yet, mean-field models rely on scale and amplitude separation between the mean and perturbation fields. The latter assumption becomes questionable for example in the nonlinear flow regimes where magnetic fluctuations, amplified by a small-scale dynamo, are no longer small compared to the mean field. Further, the assumption that conductivity variations remain small compared to the mean value is likely to fail in astrophysical plasmas where strong temperature variations are met across the flow, inducing significant variations of the magnetic diffusivity \cite{S62}. Therefore, it appears interesting to test the possibility for a dynamo to be driven {or enhanced} by conductivity variations in the general case where the hypotheses of the mean-field theory no longer hold. The present article aims at demonstrating that magnetic field generation can be enhanced by the fluctuations of the electrical conductivity of the fluid, far beyond the domain of validity of the mean-field theory, as is likely to occur in astrophysical systems. \tcb{Finally, we show that the dynamo problem with varying conductivity can be mapped onto a related dynamo problem with varying magnetic permeability. This property draws a connection between two apparently unrelated flow configurations. Using this result we justify that modulations of the magnetic diffusivity represent} a powerful asset for the design of experimental dynamos.

\section{The model}

Here we consider the kinematic dynamo problem where a magnetic field $\bf B$ is exponentially amplified by a time-independent velocity field $\bf u$ in a fluid characterised by a spatially inhomogeneous electrical conductivity $\sigma({\bf x})$. For convenience, the spatial variations of the conductivity are defined here through the variations of its inverse, the electrical resistivity $\rho({\bf x})=\sigma^{-1}({\bf x})=\rho_0 + \rho'({\bf x})$, where $\rho_0$ is the mean resistivity and $\rho'$ is the local deviation from the spatial mean. Note that in what follows, we will consider significant deviations $\rho'/\rho_0 = O(1)$ as opposed to \cite{PAG16}. For now the magnetic permeability $\mu$ is assumed to be $\mu_0$ (classical vacuum permeability).
 
The evolution of the magnetic field is governed by the induction equation, made dimensionless using the typical domain size $L$ as the unit length and the diffusive timescale $L^2\mu_0/\rho_0$ as the time unit:
\be
\label{ind}
\fb{\pd {\bf B}}{\pd t} \  =  \Rm \ \nabla \times ( {\bf u} \times {\bf B}) \ + \ \triangle {\bf B} - \ \nabla \times \big( \rho'/\rho_0 \ \nabla \times {\bf B} \big),\\
\ee
where the magnetic Reynolds number $\Rm=UL\mu_0/\rho_0$ is the control parameter quantifying the ratio between induction effects and (mean) ohmic dissipation. The magnetic field is subject to the solenoidal constraint
\be
\nabla \cdot {\bf B} \ = \ 0.
\ee
In order to investigate the effect of a variable conductivity on the dynamo threshold, we consider one of the simplest possible flow configuration for laboratory experiments: the Taylor-Vortex flow generated between two coaxial cylinders spinning at different rates. {Taylor-Vortex flow} typically builds up in a cylindrical or spherical Couette flow prone to centrifugal instability, when the angular momentum decreases outwards. It provides a simple example of incompressible, swirling flow capable of generating a dynamo magnetic field in both cylindrical  \cite{LCD00,WB02,kris14,N12} and spherical geometry \cite{MG16}. The {Taylor-Vortex} is also known to be an interesting analogue of Rayleigh-Bénard convection, which is assumed to drive many planetary and stellar dynamos \cite{B69}.

Near their onset, Taylor vortices are axisymmetric, toroidal vortices of alternating circulation and aspect ratio close to unity in the $r-z$ plane \cite{C61}. Here we consider a cylindrical domain, periodic in the axial direction, with dimensionless outer radius $r_o=2$ and vertical size $\Gamma= 4$. The inner region (with inner radius $r_i=1$) rotates as a solid body at angular velocity $\Omega_i=1$. The region $r_i < r \le r_o$ corresponds to the fluid domain, where the velocity field prescribed here essentially reproduces the basic features of the saturated {Taylor-Vortex} pattern. The azimuthal velocity simply reduces to the classical Couette solution for a viscous flow entrained by two infinite, coaxial cylinders with the outer cylinder at rest ($\Omega_o=0$):
\begin{align}
{\bf u} \cdot {\bf e_{\phi}} &= \Omega_i r   &  \text{for } r \le r_i,\\
 {\bf u} \cdot {\bf  e_{\phi}} &= \fb{\Omega_i}{r_i^{-2}-r_o^{-2}} (r^{-1}-r_o^{-2}r)  & \text{for }  r_i < r \le r_o.
 \label{couette}
\end{align}
The prescribed radial and axial velocities define two pairs of axisymmetric, counter-rotating toroidal vortices:
\be
{\bf u} \cdot {\bf e_{r}}= \pd_z \Psi \qquad \text{and} \qquad {\bf u} \cdot {\bf e_{z}}= - \pd_r \Psi,
\ee
where the streamfunction $\Psi$ is defined here as
\begin{align}
\label{psi}
\Psi&= 0   &\text{for } r \le r_i,\\
 \Psi &=  6(r-r_i)^2 (r_o-r)^2 \cos(\pi z)  &\text{for } r_i < r \le r_o.
\end{align}
Our use of a largely simplified flow to represent the saturated {Taylor-Vortex flow} evidently implies that the kinematic dynamo threshold is expected to differ quantitatively from that of a true, laminar Taylor-Couette flow with the same domain aspect-ratio. Indeed it is important to emphasize that our purpose here is \textit{not} to quantitatively reproduce full magnetohydrodynamic (MHD) simulations, but rather to show the effect of inhomogeneous fluid properties on the dynamo threshold in a simple but realistic flow configuration, where variations of both $\Psi$ and $\rho'$ are explicitly formulated.

Finally, the prescribed electrical resistivity deviation $\rho'$ is axisymmetric and exhibits periodic variations along the axial direction:
\begin{align}
\rho'/\rho_0 &= 0   &\text{for } r \le r_i,\\
\label{eta}
\rho'/\rho_0&= {4 \lambda} \ (r-r_i) (r_o-r) \cos(2 \pi z)  &\text{for } r > r_i.
\end{align}
where $\lambda$ is a tunable coefficient $0 \le \lambda < 1$ describing the intensity of the spatial variations. The choice for the particular resistivity modulation (\ref{eta}) will be made clear in the next section. Note that the resistivity modulation prescribed here has zero mean, so that a possible modification of the dynamo threshold cannot be related to a modification of the effective magnetic Reynolds number.

\begin{figure*}
\begin{center}
\hspace{-5ex}
\includegraphics[width=\textwidth,clip=true]{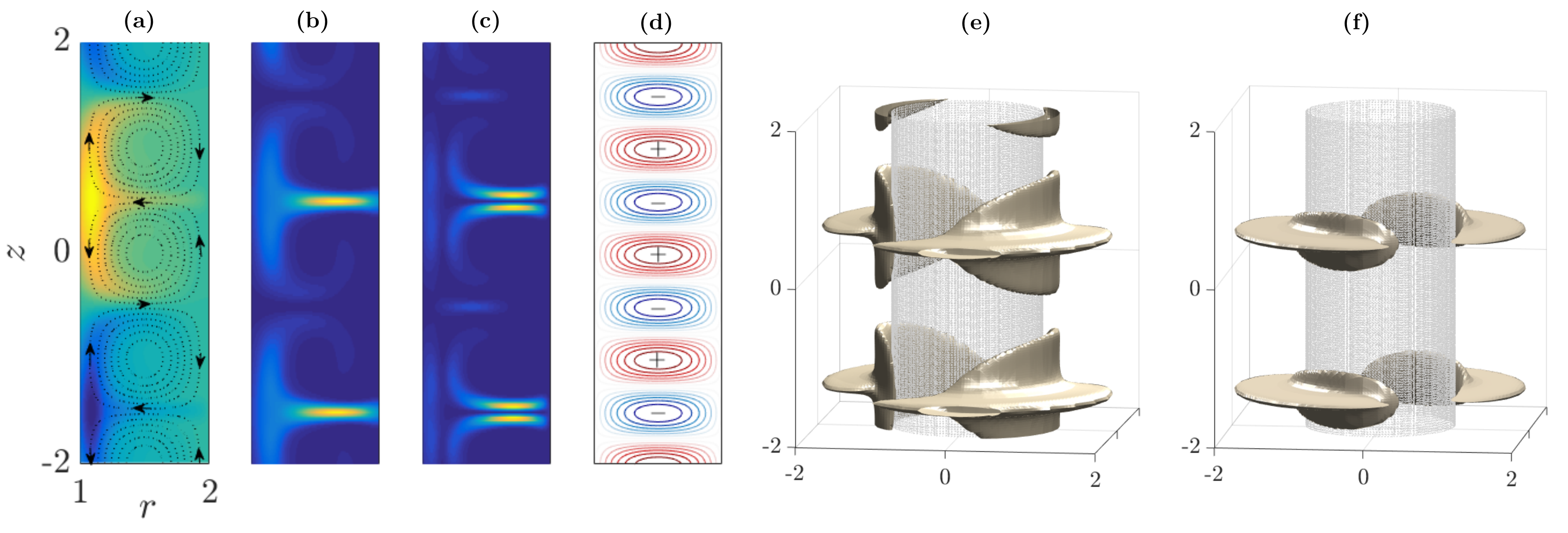}
\caption{\small{\textit{From Left to Right:} Axial slices of (a) the azimuthal magnetic field in the fluid domain (colormap), with streamline pattern superimposed ($\Psi$ isocontours); (b) magnetic energy density (averaged in the azimuthal direction; (c) electrical energy density (again, averaged in the azimuthal direction); all of them corresponding to the reference case $\Rm=100$, $\lambda=0$. (d): Isocontours of the eletrical resistivity modulation $\rho'/\rho_0$ as defined in (\ref{eta}). (e)-(f): Isosurfaces of magnetic energy (drawn for $10\%$ of the maximum value), (e) $\lambda=0$, $\Rm=100$; (f) $\lambda=0.9$, $\Rm=80$. The inner cylinder boundary is shown for reference.}}
\label{fig:maps}
\end{center}
\end{figure*}

The induction equation (\ref{ind}) with variable electrical resistivity is discretized in space using a second order, finite differences method on a cylindrical, staggered mesh \cite{Y66} for preservation of the magnetic field divergence. The numerical scheme used here for time-integration is a second-order semi-implicit backward differentiation scheme, where the diffusion operator is dealt with implicitly, whereas the induction operator is treated explicitly and linearly extrapolated to the upcoming timestep \cite{A95}. 

Ferromagnetic (pseudo-vacuum) boundary conditions (infinite-permeability boundary yielding ${\bf B \times n = 0}$, where $  n$ is the normal vector to the boundary) are prescribed on the outer cylinder $r=r_o$. The domain is periodic in the axial direction. 
The typical resolution used for the direct numerical simulations presented in the next section is $128 \times 32 \times 128$ gridpoints, which convergence tests (up to $192 \times 64 \times 256$) have shown to be sufficient for the purpose of the present study.

\section{Results}

When no spatial variations of electrical resistivity are included in our model ($\lambda=0$), a non-axisymmetric, non-purely toroidal magnetic field perturbation eventually undergoes exponential amplification if the magnetic Reynolds number exceeds the critical value $\Rm_c \sim 73.5$. Note that $\Rm_c$ depends on both the flow structure and (importantly) the chosen magnetic boundary conditions \cite{G08a}. As a result, the value of $\Rm_c$ found here with our simplified flow model cannot be quantitatively compared with the kinematic dynamo threshold $\Rm_c=134.9$ found in the same geometry by \cite{LCD00} for Taylor Couette flow using different boundary conditions. Nevertheless, the first magnetic eigenmode ${\bf B}(r)e^{i(m_B \phi + k_B z)}$ is characterised by $m_B=1$ in the azimuthal direction and $k_B=\tfrac12 k_V$ in the axial direction (where $k_V=\pi$ describes the velocity field axial mode), thus displaying the same magnetic field morphology as the kinematic dynamo in \cite{LCD00}. Importantly, it also displays the same magnetic field morphology as the fully nonlinear dynamos found by \cite{kris14,WB02} for \tcb{realistic,} axisymmetric {Taylor-Vortex flow} using the same domain geometry. The structure of the first dynamo mode is illustrated in Figure \ref{fig:maps}a-c, where an axial slice of the azimuthal magnetic field is shown along with colormaps of the (azimuthally averaged) magnetic and electrical energies. Streamlines are superimposed for reference, showing that magnetic energy typically builds up in the stagnation region between two counter-rotating vortices and in the vicinity of the inner cylinder boundary, where differential rotation is stronger.

In what follows, the electrical resistivity is now spatially modulated in the axial direction. In the context of small-amplitude perturbations, \cite{PAG16} \tcb{has} shown that the mean-field effect due to conductivity fluctuations (the $\alpha^\sigma$-effect) operates when non-diagonal coefficients of the $\alpha^\sigma$ tensor are non-zero. In practice, this corresponds to the requirement that flow vorticity gradients coincide with $\sigma$ extrema (or conversely, that $\sigma$ gradients coincide with vorticity extrema). A following paper has shown a similar effect for anisotropic but spatially homogeneous conductivity: in this case, the magnetic diffusivity $\eta$ is no longer a scalar field but a tensor, whose non-diagonal coefficients have to be non-zero to observe a reduction of the dynamo threshold \cite{PA20}.
Bearing these results in mind, it would be natural to define the electrical resistivity deviation $\rho'/\rho_0$ as a sinusoidal function of $z$ such that its variations are shifted by $\pi/2$ compared to the streamfunction $\Psi$. Yet, the distribution of electrical currents in the reference dynamo field (Figure \ref{fig:maps} with $\lambda=0$) suggests that a resistivity modulation such as (\ref{eta}), at twice the wavelength of $\Psi$ in the axial direction, should provide an even more efficient $\rho'$-dynamo, as it would also enhance the secondary peaks in current intensity (observable here at  $z \sim -0.5$ and $z \sim 1.5$ in Figure \ref{fig:maps}c). Such a modulation corresponds to the case where the electrical resistivity becomes minimal (or conversely, where the conductivity becomes maximal) for each zero of the vorticity field, regardless of the sign of the vorticity variation (see Figure \ref{fig:maps}d).

\begin{figure}
\begin{center}
\includegraphics[width=0.5\textwidth,clip=true]{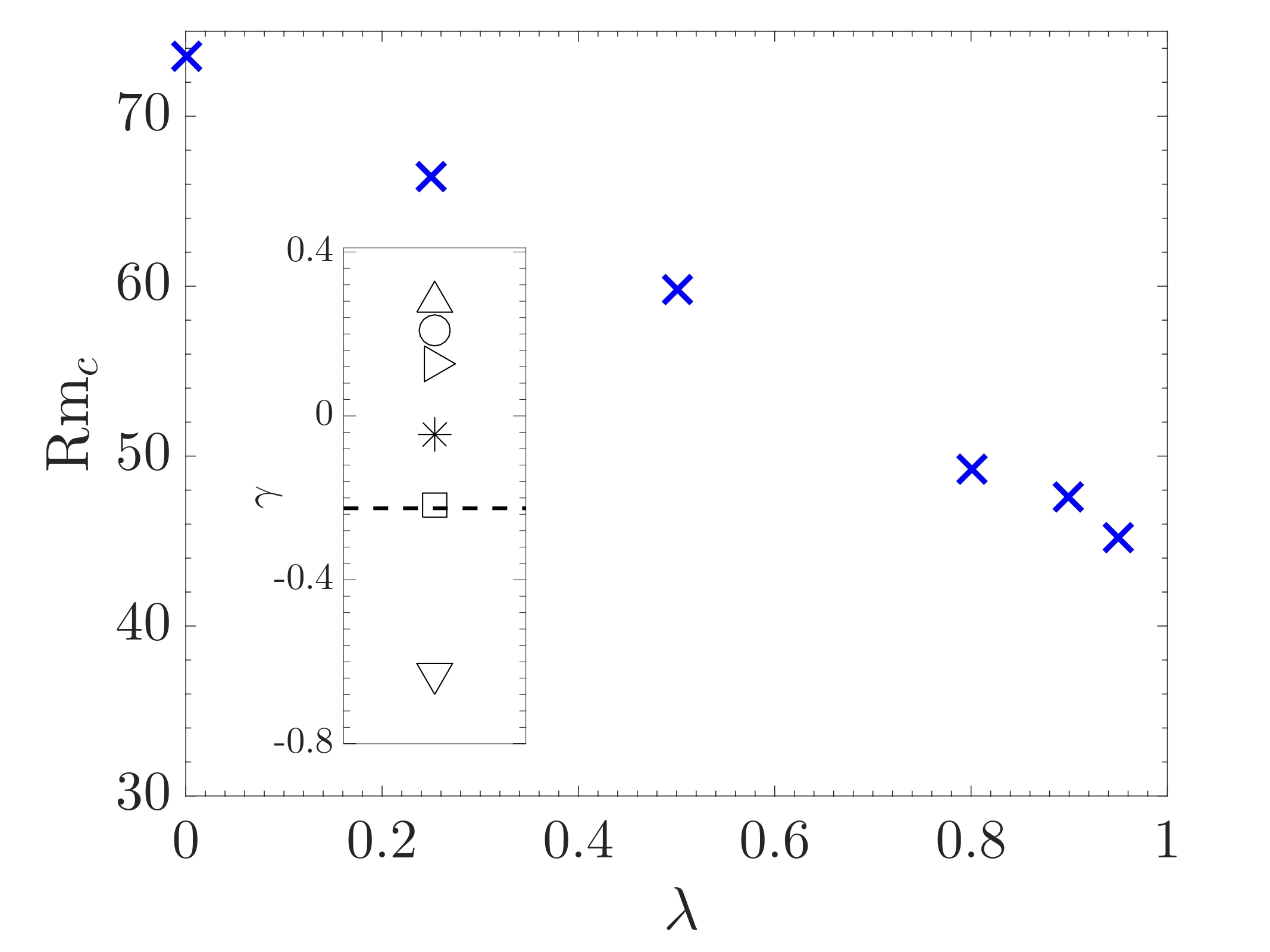}
\caption{\small{Critical magnetic Reynolds number for kinematic dynamo $\Rm_c$ as a function of the amplitude of the resistivity modulation $\lambda$. Note that the prefactor in (\ref{eta}) ensures that, for example, $\lambda=0.5$ corresponds to a \tcb{maximum} resistivity variation of $50\%$ around its mean value. {\textit{Inset:} Asymptotic kinematic growth rate for $\Rm=70$ and $\lambda=0.25$, with different values of the phase shift $\Phi$ in (\ref{shift}): $\Phi=\pi \, (\triangledown)$; $\pi/2 \, (\square)$;  $\pi/4 \, (\triangleright)$; $\pi/6 \, (\circ)$ and $0 \, (\vartriangle)$. The symbol $\ast$ corresponds to a different modulation wavelength with $\rho'\propto \cos(\pi z+\pi/2)$, and the dashed line marks the value of the reference growth rate ($\lambda=0$).}}}
\label{fig:Rmc}
\end{center}
\end{figure}

The resistivity modulation (\ref{eta}) induces no structural changes in the dominant magnetic mode, as shown in Figure \ref{fig:maps}e-f , where isosurfaces of the magnetic energy density are drawn respectively for $\lambda=0$ (reference case) and $\lambda=0.9$. The chosen conductivity modulation merely tends to concentrate the magnetic energy in the regions ($z \sim -1.5$ and $z \sim 0.5$) where the stretching of magnetic field lines is strongest, due to the combined effects of strong differential rotation in the azimuthal direction (near the inner cylinder boundary) and diverging flow pattern (near the stagnation point between two counter-rotating vortices). 
Yet, the effect of the resistivity modulation on dynamo onset is remarkable:  Figure \ref{fig:Rmc} shows the critical magnetic Reynolds number $\Rm_c$ for kinematic dynamo action found for spatial variations of increasing amplitude $\lambda$:  for the largest-amplitude modulation presented here ($\lambda=0.95$), the dynamo threshold is reduced by $38\%$ compared to its reference value ($\lambda=0$ case). {Note that (\ref{eta}) has been tailored here so as to efficiently lower the dynamo threshold. The opposite can evidently occur with a different modulation, in particular when a different phase shift between the vorticity and the resistivity pattern is chosen. This is illustrated by the inset of Figure \ref{fig:Rmc} where (\ref{eta}) has been replaced by the modulation:}
\be
\label{shift}
{\rho'/\rho_0= {4 \lambda} \ (r-r_i) (r_o-r) \cos(2 \pi z + \Phi),}
\ee
{and the kinematic growth rate is shown for different values of the phase $\Phi$, with fixed $\lambda$ and $\Rm$.}

\section{Discussion}

Using duality arguments, \cite{FP13} showed that, up to a change of sign in the velocity field, the problems of kinematic dynamo growth in a fluid domain with either infinite electrical conductivity (``perfect conductor'') or infinite magnetic permeability (``ferromagnetic'') boundaries are equivalent. In point of fact, this observation is extendable to the duality between the problems of inhomogeneous electrical conductivity on the one hand, and inhomogeneous magnetic permeability on the other hand. Indeed the induction equation for the potential vector $\bf A$ (with $\bf B = \nabla \times A$) reads in the general case where both the electrical conductivity $\sigma$ and the magnetic permeability $\mu$ are spatially inhomogeneous:
\be
\label{indA}
\fb{\pd {\bf A}}{\pd t} \ = \ \mathscr{L} \B  \ \equiv \ \u \times \B - \fb{1}{\sigma({\bf x})} \nabla \times \left( \fb{\B}{\mu ({\bf x})} \right),
\ee
where the Weyl gauge has been used. For any magnetic potential vector $\bf A$ such that $\bf A \times n = 0$ on the (impermeable) domain boundary $\pd \Omega$, and for any magnetic field $\Bt$ such that $\Bt \times {\bf n} = 0$ on $\pd \Omega$,
\be
\lel \Bt \ , \ \mathscr{L} \, \nabla \times {\bf A} \rir =  \lel \nabla \times \mathscr{L}^* \Bt \ , \ {\bf A} \rir
\ee
with the inner product $\lel {\bf a \ , b} \ \rir = \int_\Omega {\bf a \cdot b} \, d\Omega$, and the linear operator $\mathscr{L}^* \Bt \equiv (- \u) \times \Bt - \fb{1}{\mu({\bf x})} \nabla \times \left( \fb{\Bt}{\sigma ({\bf x})} \right)$.

Because the eigenvalues of the induction operator (acting on \textit{potential vectors}) $\mathscr{L} \nabla \times $ are the complex conjugates of that of its adjoint operator $\nabla \times \mathscr{L}^*$ (acting on \textit{magnetic fields}), the kinematic induction problem (\ref{indA}) with infinitely conducting boundaries (as ensured by $\bf A \times n = 0$ on $\pd \Omega$) meets the same dynamo threshold as the dual problem where ferromagnetic boundaries ($\Bt \times \bf n = 0$ on $\pd \Omega$) and the reverse velocity field ${- \u}$ are considered, provided $\sigma({\bf x})$ and $\mu({\bf x})$ are swapped in the governing equation for the magnetic field:
\be
\fb{\pd \Bt}{\pd t} \ = \  \nabla \times  (-\u \times \Bt) - \nabla \times \left( \fb{1}{\mu({\bf x})} \nabla \times \left( \fb{\Bt}{\sigma ({\bf x})} \right)\right).
\ee
The Taylor-Vortex flow is left invariant by the sign-reversal transformation $u \rightarrow -u$, combined with (i) mirror symmetry with respect to any plane containing the axis of rotation and (ii) translation by half a period along the axial direction. \tcb{As a consequence,} the numerical results presented in the previous section (with ferromagnetic outer cylinder and inhomogeneous electrical conductivity) carry over to the equivalent setup where the magnetic permeability is inhomogeneous instead of the electrical conductivity, and the outer cylinder is a perfect conductor. 

Beyond the particular {Taylor-Vortex flow}, the duality arguments developed above can be extended to describe fluids and boundaries with arbitrary electrical and magnetic properties. To wit, the fields are defined over $\reals^3$ and the distributions $\mu({\bf x})$ and $\sigma({\bf x})$ describe the local electrical and magnetic properties both inside the fluid domain and outside of it (i.e. inside the boundaries). One readily concludes that the dual problem is obtained by substituting $u \rightarrow -u$ and swapping $\mu({\bf x})$ and $\sigma({\bf x})$ both \textit{inside} and \textit{outside} the fluid. As a special case, we recover the situation described above when the boundary is perfectly insulating, $\sigma \to \infty$ in the boundary, which is dual to the case of a ferromagnetic boundary with $\mu \to \infty$. But the result holds for any finite $\mu({\bf x})$ and $\sigma({\bf x})$, be them uniform or not. For instance, solid boundaries with either inhomogeneous magnetic permeability~\cite{GPF12,GPF13} or inhomogeneous electrical conductivity~\cite{BW92} were shown independently to be sources of dynamo action. The analysis above indicates that these two dynamo setups are also dual problems that can be mapped onto one another. \tcb{As a consequence, for each mode of the first problem with growthrate $p_i$, there exists a different mode of the second problem with growthrate $p_i^*$, implying in particular that the onsets of instability are equal for both setups.}

We now discuss the possibility to observe this effect in a laboratory experiment. Relying on full MHD simulations, \cite{kris14} showed that the onset of the (homogeneous) {Taylor-Vortex} dynamo is strongly decreased by inserting equally spaced blades separated in the axial direction by a distance equal to the width of the fluid domain, and alternatively attached to the inner and outer cylinder. In this configuration, the threshold decreases from $Rm_c \approx 140$ to $Rm_c \approx 80$, a value similar to the one obtained for our synthetic flow in the homogeneous case ($\lambda=0$). This decrease is easy to understand: the full MHD simulations presented in \cite{kris14} for freely evolving {Taylor-Vortex flow} (without blades) show radial velocities of order $0.2$ (let us recall that the velocity unit is set by the azimuthal velocity on the inner cylinder), whereas introducing blades induces $O(1)$ azimuthal and radial velocities in the {von K\'arm\'an} and B\"odewadt layers attached to the blades upper and lower boundaries. Hence, the additional fluid entrainement supplied by the blades seems to contribute in multiple ways in lowering the dynamo threshold: not only do the blades ``channel'' the Taylor vortices and thus preserve coherent mean flow further into the turbulent regime - the Taylor-Couette dynamo onset being very sensitive to the coherence of the vortex structure \cite{kris14,MG16} - but they also tend to increase significantly the ratio between toroidal and poloidal velocities, which has been shown to favor {\tcb{Taylor}-Vortex} dynamo action \cite{WB02}.
Interestingly, the simple conductivity modulation described in the previous section precisely corresponds to this configuration: the blades, having material properties different from that of the stirred fluid, will also locally provide a very favorable resistivity modulation, as the minima of $\rho$ then coincide with the interfaces between consecutive Taylor vortices. {The bladed configuration described above could be implemented in the laboratory} using soft iron blades (relative permeability $\mu_r=100$) and liquid sodium ($\mu_r=1$, $\sigma=10^{7}$ $\Omega^{-1}$m$^{-1}$), which is analogous to a resistivity modulation $\lambda$ of order one. In this case, our simulations predict that the onset should be further decreased from $Rm_c\sim 80$ to $Rm_c\sim 45$: for a cylinder of inner radius $r_i=15$cm, outer radius $r_o=30$cm and height $60$cm, this leads to a critical rotation rate $\Omega_c=240$rad.$s^{-1}$. Note that this reduction of $\Omega_c$ by a factor $3$ compared to the non-modified {Taylor-Vortex} dynamo  ($\Omega_c \sim 750$ rad.$s^{-1}$) translates into a factor $30$ for the energy injection rate: indeed, the rotation rates at stake here ensure that the Taylor-Couette flow lies well in the inertial regime where the energy dissipation rate is proportional to $\Omega^3$ \cite{EGL07,GLS16}. 
 The modulation of magnetic diffusivity through the insertion of metallic blades with high magnetic permeability could therefore provide a very promising way of achieving a Taylor-Couette dynamo in the lab. 

This mechanism may also provide a simple explanation for {some} results of the VKS laboratory dynamo experiment, in which two co-axial counter-rotating bladed discs drive a {von} K\'arm\'an swirling flow of liquid sodium. Such a flow is prone to dynamo action at sufficiently large Rm, but this was observed only when impellers made of soft iron (i.e. ferromagnetic impellers) were used \cite{M07}. {While it has been shown numerically that increasing the magnetic permeability of the impellers does indeed lower the dynamo threshold (see e.g. \cite{Giesecke2010,Johann14,Yannick2017} and references herein), it is interesting to note that the flow ejected by the centrifugal force in the region close to the impellers is strongly helical due to the generation of trailing vortices behind each of the ferromagnetic disk blades \cite{PMF07,kris09}. In other words, it is remarkable that the threshold of the VKS dynamo seems lowered when the vorticity pattern presents a strong correlation with the distribution of magnetic permeability, which is reminiscent of the mechanism described in the present paper.} 

The problem described here could also have interesting astrophysical applications. For instance, the transition to turbulence in accretion disks remains a puzzling mystery, and one of the most promising explanation involves a nonlinear coupling between a dynamo and a magnetorotational instability \cite{B03,B11}, a problem which is deeply connected to the destabilisation of quasi-Keplerian Taylor-Couette flows \cite{Johann11,Gal17}. Because the ionized gas of accretion discs presents significant variations in density, taking into account these density-induced resistivity fluctuations could fundamentally modify the question of magnetic field generation in these objects.

\paragraph{Acknowledgements} -- FM acknowledges support from the French program `T-ERC' managed by Agence Nationale de la Recherche (Grant ANR-19-ERC7-0008-01). 
CG acknowledges support from the French program `JCJC' managed by Agence Nationale de la Recherche (Grant ANR-19-CE30-0025-01). FP acknowledges support from Agence Nationale de la Recherche (Grant \tcb{ANR-19-CE31-0019-01}).

\end{document}